\documentclass[10pt,aps,pra,twocolumn,showpacs,superscriptaddress,nobalancelastpage,notitlepage]{revtex4-1}

\usepackage{graphicx,psfrag,amsmath,amssymb,amsfonts,bbm,latexsym,color,dcolumn,bm,xcolor}

\begin{document}

\title{Black hole production at lepton colliders}

\author{Hang Qi}

\affiliation{\mbox{Department of Physics and Astronomy, Dartmouth College, 6127 Wilder Laboratory, 
Hanover, NH 03755, USA}}

\author{Roberto Onofrio}

\affiliation{\mbox{Dipartimento di Fisica e Astronomia ``Galileo Galilei'', Universit\`a  di Padova, 
Via Marzolo 8, Padova 35131, Italy}}

\affiliation{\mbox{Department of Physics and Astronomy, Dartmouth College, 6127 Wilder Laboratory, 
Hanover, NH 03755, USA}}

\begin{abstract}
Production of black holes has been discussed in a variety of extensions of the Standard 
Model, and related bounds have been established from data taken at the Large Hadron Collider. 
We show that, if the Higgs particle has a fully gravitational content via the equivalence principle, enhanced cross-sections of black holes at colliders should be expected within the Standard Model itself. The case of black hole production by precision measurements at electron colliders is discussed. The Coulomb repulsion strongly suppresses the related cross-section with respect to the one based on the hoop conjecture, making the possible production of black holes still unfeasible with current beam technology. At the same time, this suggests the reanalysis of the bounds, based on the hoop conjecture, already determined in hadronic collisions for extra-dimensional models. 
\end{abstract}

\maketitle

\section{Introduction}

The possibility that the Fermi scale and the Planck scale actually coincide is an appealing feature of many extensions of the Standard Model as it avoids the hierarchy problem. This may  occur either by embedding the latter in higher-dimensional space-time \cite{Arkani,Antoniadis,Randall}, by introducing a hidden sector with a large number of fermions \cite{Dvali1,Calmet}, or invoking spin-related frame-dragging effects \cite{Burinskii}. In these scenarios the effective gravitational coupling constant at the Fermi scale should be then increased by about 33 orders of magnitude with respect to the one measured in the macroscopic world.  In this way, its value becomes identical or of the same order as the coupling constants involved in charged weak interactions via the Fermi constant $G_F$ and the CKM and PMNS flavor mixing matrices. As such, gravitational effects should be already observable at the Fermi scale, and quite prominent among these should be the formation of black holes in high-energy, small 
impact parameter collisions. The hoop conjecture \cite{Thorne} assumes that production of black holes should occur with significant probability whenever the distance between the two colliding particles becomes comparable to their Schwarzschild radii. Semiclassical considerations suggest that, for collisions at center of mass energy $\sqrt{s}$, the 
cross-section for black hole production should be of the order  $\sigma_{ BH} \sim \pi \tilde{R}_S^2$ \cite{Giddings,Dimopoulos}. 
Here $\tilde{R}_S=2 \tilde{G}_N \sqrt{s}/c^2$  is the Schwarzschild radius, and $\tilde{G}_N$ is the effective gravitational Newton constant 
at the Fermi scale, whose numerical value is assumed to be much larger than the known gravitational Newton constant $G_N$ measured at the macroscopic scale. A further intriguing implication is that high-energy physics with production of new particle states should be inhibited \cite{Giddings}, leading to a dominance of gravitational physics over quantum field theory beyond the Fermi scale, in some cases implying a revival of classicality \cite{Dvali2}.
 
Searches for black holes, at the Large Hadron Collider (LHC) general purpose detectors CMS \cite{CMSBH1,CMSBH2} 
and ATLAS \cite{ATLASBH1,ATLASBH2}, have placed first lower bounds on their mass in the 5-10 TeV range. 
However, these limits are strongly model dependent, especially in regard to the variety of assumptions made on the decay products 
of the black holes. In particular, criticisms have been raised about the validity of the Hawking radiation emission mechanism during the black hole decay process. The Hawking effect has been derived in a semiclassical regime in which the mass of the black hole is considered to be much larger than the putative lower Planck mass scale, and this is in conflict with the assumptions made in the data analysis \cite{Park}.
The large background due to hadronic interactions and the consequent large multiplicity of events, and the presence of  
initial states that cannot be well controlled at hadron colliders, are also far from optimal features from the standpoint of black 
hole production and characterization. 

In this letter, we focus on black hole production as a mechanism intrinsic to the Standard Model, and we sketch a proposal for its test by using electron colliders. This 
analysis complements various former contributions more focused on black hole decay modes. 
The leptonic nature of the colliding particles lowers the overall rate of uninteresting, from this perspective, hadronic events. Using same-charge particles avoids the usual annihilation $s$-channel otherwise allowed for particles-antiparticles collisions \cite{Barkas,Barber}. The price payed in looking for black hole production in electron colliders is the presence of Coulomb repulsion, which modifies significantly the estimates based on the hoop conjecture.

\section{Higgs-related gravity at the Fermi scale}

The discovery of a scalar resonance compatible with the expectations for a Higgs particle \cite{ATLASHiggs,CMSHiggs} has consolidated the 
Standard Model by opening up the possibility of experimentally testing electroweak symmetry breaking and a 
new interaction, that of the Higgs particle with all other fundamental particles. 
A unique feature of the Higgs-fermion coupling is the fact that its strength is proportional to the inertial mass of the fermions. 
If the equivalence principle holds at the microscopic scale, this would imply that the Higgs should also couple 
linearly to the gravitational mass of the fermions, and this closely mimics what is expected for  gravitational couplings in the nonrelativistic limit. 
With respect to the gravity we experience in the macroscopic world, the Higgs-related gravity should be short range 
(with a Yukawa range on order of the Compton wavelength associated to the Higgs particle), with  coupling 33 orders of 
magnitude larger than macroscopic gravity, and scalar, rather than tensorial, in character. 
This possibility has been discussed in detail in \cite{Dehnen1,Dehnen2}. However, although evaluating at the very end 
of \cite{Dehnen1} the magnitude of the expected interaction, the authors have not made connection to the weak 
interaction coupling strength. A comparative discussion of the similarities between gravitational and weak interactions 
appeared in \cite{Onofrio1}, including the discussion of potential phenomenological implications. 
In particular, these include the possible impact in precision spectroscopy of muonic hydrogen \cite{Onofrio2}, a gravitational 
interpretation of the Yukawa coefficients, and the impossibility of observing gauge-mediated particles at energies larger than 
the Higgs vacuum expectation value \cite{Onofrio3}, confirming the prediction already discussed in \cite{Giddings}. 
From this perspective, the gravitational content of the Higgs field is similar to that of the scalar component in the 
Brans-Dicke theory of gravitation \cite{Brans}. In the following, we will consider an 
effective gravitational potential sourced by the usual infinite-range Newtonian term, of tensorial origin in general relativity, as 
well as from a short-range Higgs term of scalar origin, both directly proportional to the mass of the source $m$, in the form 

\begin{equation}
V_{\mathrm{eff}}(r)= -\frac{G_N m}{r}\left(1+\alpha_H e^{-r/\lambda_H}\right),
\label{Yukawa}
\end{equation}
with $\alpha_H=1.23 \times 10^{33}$ and $\lambda_H=h/(m_H c) \simeq 10^{-17}$ m the Compton wavelength of the Higgs  \cite{Onofrio2}. At distances much larger than $\lambda_H$  this potential corresponds to Newtonian gravity, but at distances of the order of $\lambda_H$ or smaller, this potentially boosts the effective gravitational coupling by the factor $\alpha_H$. 
This parameterization of strong gravity effects is rather simple with respect to the case of extra-dimensional physics, it has similar quantitative predictions, and allows for a simpler comparison to the other interactions present in the case of charged leptons, {\it i.e.} the electromagnetic interaction.
Considering this effective interaction and the repulsive Coulomb interaction allows for the exploration of the actual dynamics of collisional processes, as well as the related predictions for the probability to form black holes, as we describe in the next sections.

\begin{figure*}[t]
\includegraphics[width=0.25\textwidth, clip=true]{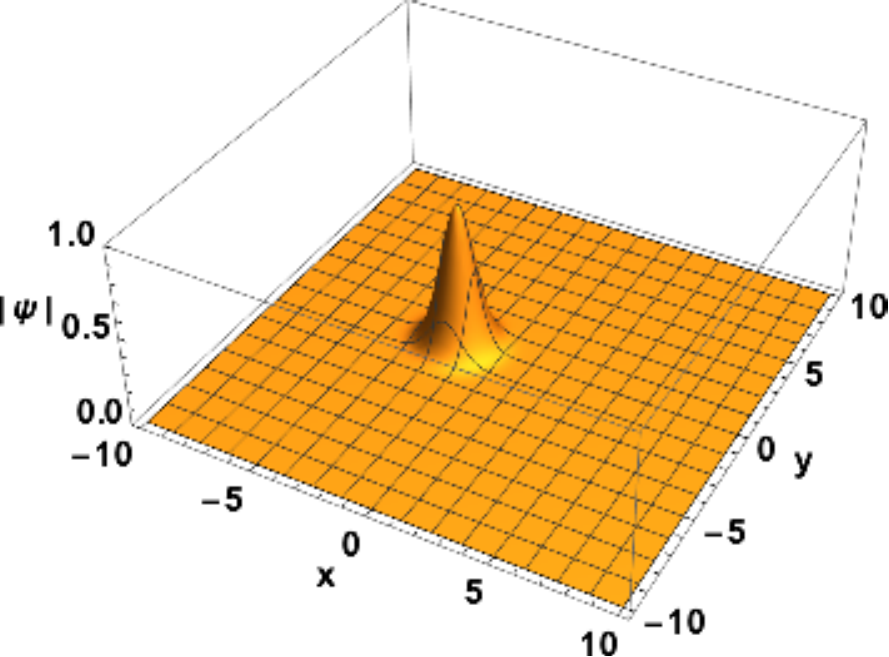}
\includegraphics[width=0.25\textwidth, clip=true]{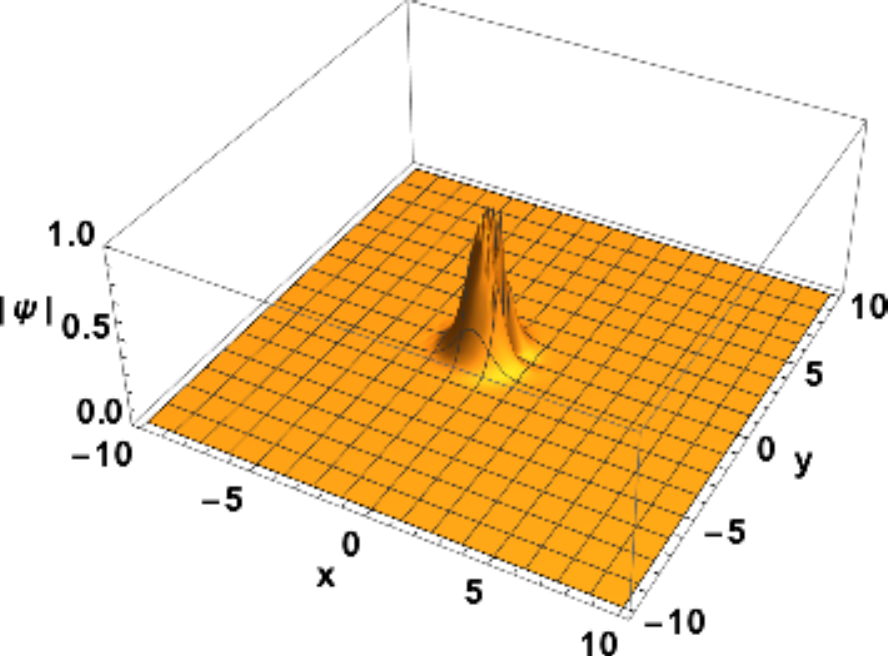}
\includegraphics[width=0.25\textwidth, clip=true]{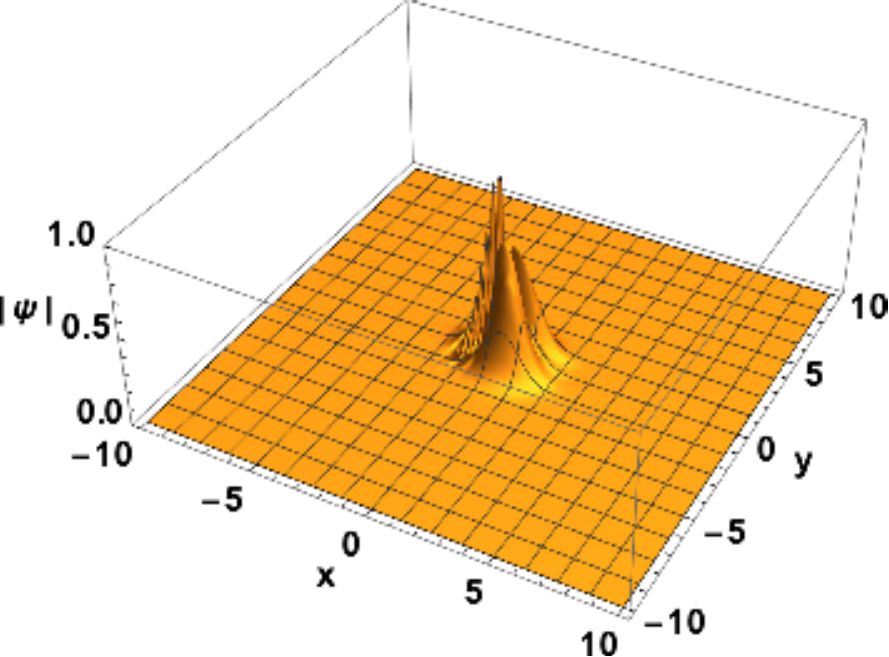}
\includegraphics[width=0.25\textwidth, clip=true]{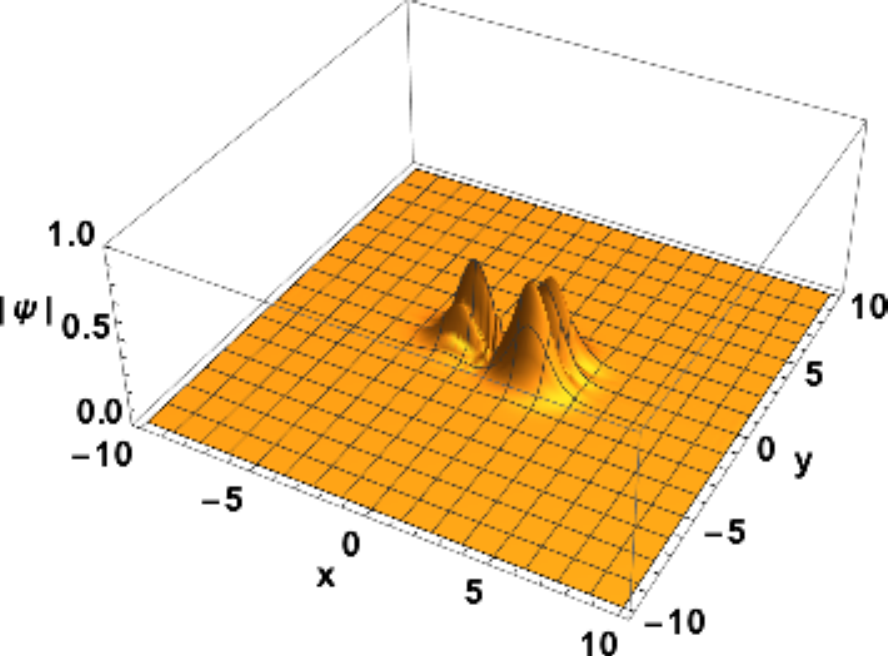}
\includegraphics[width=0.25\textwidth, clip=true]{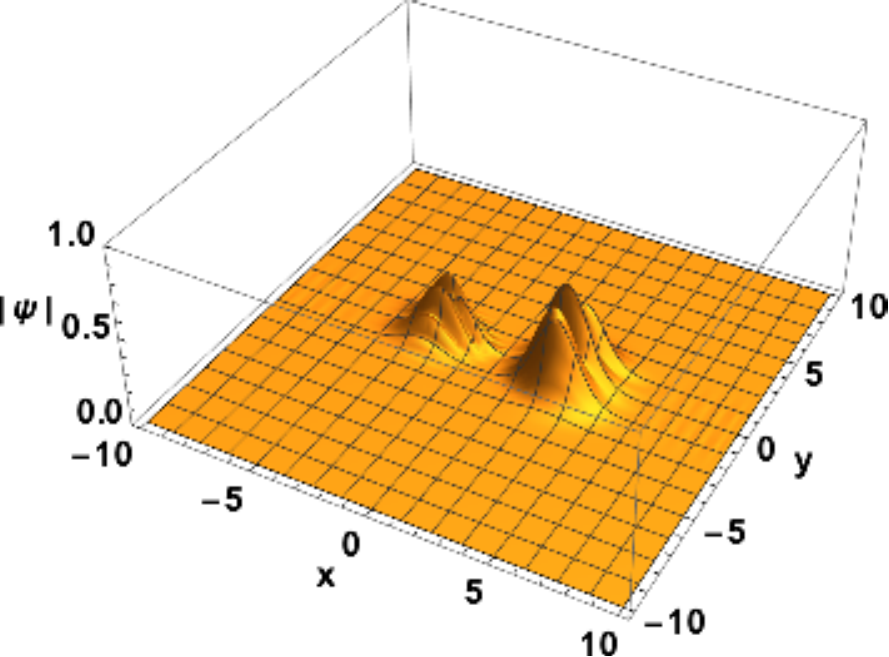}
\includegraphics[width=0.25\textwidth, clip=true]{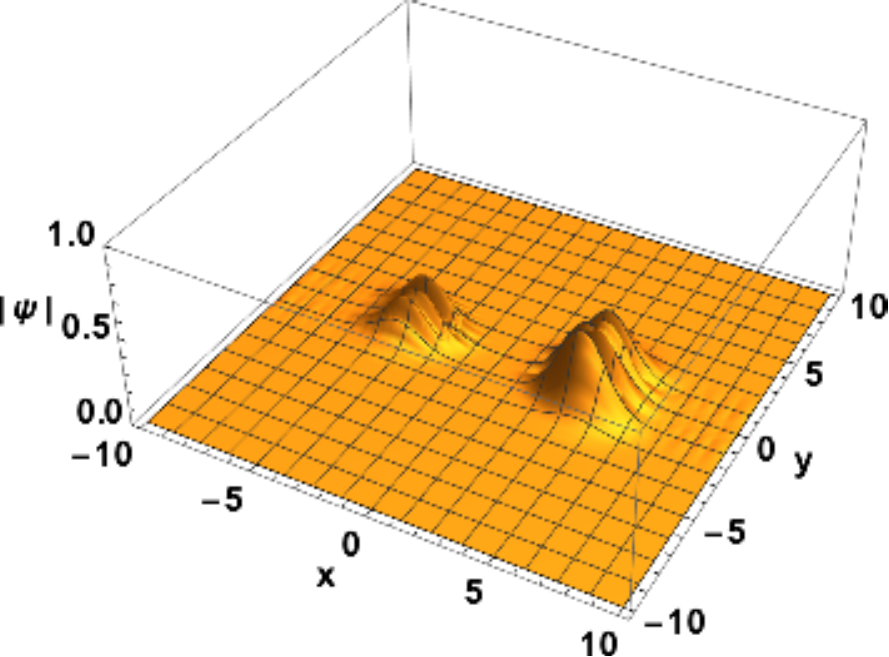}
\caption{Snapshots of the modulus of the wave function for a wave packet traveling with a given initial velocity and impact parameter in the presence of a potential 
 with a source located in the center, evaluated through numerical simulations of the Schroedinger equation with a purely repulsive Coulomb potential. The case shown is relative to a nearly head-on collision, with an impact parameter of 0.1 in spatial units, and a momentum of 0.15 in reciprocal spatial units.   The snapshots are taken at times (in arbitrary units, and from top left to bottom right) t=0, 0.05, 0.1, 0.15, 0.2, 0.25.}
\label{Fig1}
\end{figure*}

\section{Electron-electron elastic scattering}

The increase in black hole formation from the enhanced gravity due to the Higgs 
field, or any mechanism producing strong  gravity at the microscopic scale, is given by estimating the 
Schwarzschild radius. For ordinary gravity and the case of the electron, the Schwarzschild radius is 
$R_S=2 G_N m_e/c^2=1.3 \times 10^{-57}$ m, and by boosting the Newton constant to the value 
$\tilde{G}_N=1.23 \times 10^{33} G_N$ we get $\tilde{R}_S=1.6 \times 10^{-24}$ m. This means a hoop-conjecture-based cross-section with a peak value 
of $\sigma_{BH} \sim 8 \times 10^{-48} {\mathrm{m}}^2 \sim 8 \times 10^{-5}$ fb for impact parameters much smaller than the Higgs Compton wavelength $\lambda_H$. In the intermediate case, we will interpolate the effective Schwarzschild radius as 
$\tilde{R}_S(r) = R_S [1+\alpha_H \exp(-r/\lambda_H)]$. 
This is based on the assumption that, due to the Higgs field, even the scalar component of the potential contributes to the space-time metric and to the emergence of the event horizon. A detailed discussion of black holes collapse in Brans-Dicke theory makes this hypothesis plausible \cite{Hawking,Scheel1,Scheel2,Sotiriou}.
Notice also, as further elements of plausibility, that the Yukawa component does not qualitatively change the functional behavior of the potential at any length scale. Furthermore, at the length scale at which the horizon is expected (not larger than $10^{-24}$ m), the Yukawa term of the Higgs 
potential is, for all practical purposes, a $1/r$ potential like standard gravity but with a larger coupling constant.
In the case of relativistic electrons with energy  $E$  the Schwarzschild radius is further increased by the 
relativistic factor $\gamma=E/(m_e c^2)$. For instance, for electron-electron collisions occurring at 100 GeV+100 GeV, we have 
$\gamma \simeq 2 \times 10^5$, and the peak value of the black hole cross-section 
will be boosted to a value of about 3.2 nb. 

These estimates based on purely geometric considerations are quite optimistic as they do not take into account that 
colliding particles interact. More realistic estimates of the black hole production cross-section are therefore obtained 
by investigating the full collision process. In doing this we try to single out the basic effects neglecting various factors 
which are not expected, at leading order, to change significantly the estimates. First, we neglect the effect of spin 
in the collisions, therefore schematizing the electrons as usually done in scalar electrodynamics.
\begin{figure*}[t]
\includegraphics[width=0.25\textwidth, clip=true]{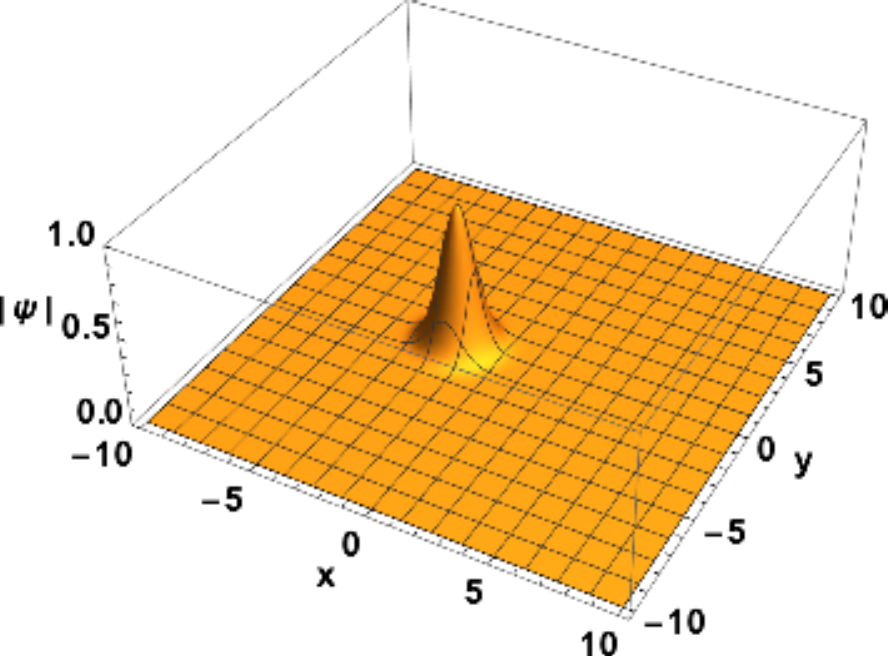}
\includegraphics[width=0.25\textwidth, clip=true]{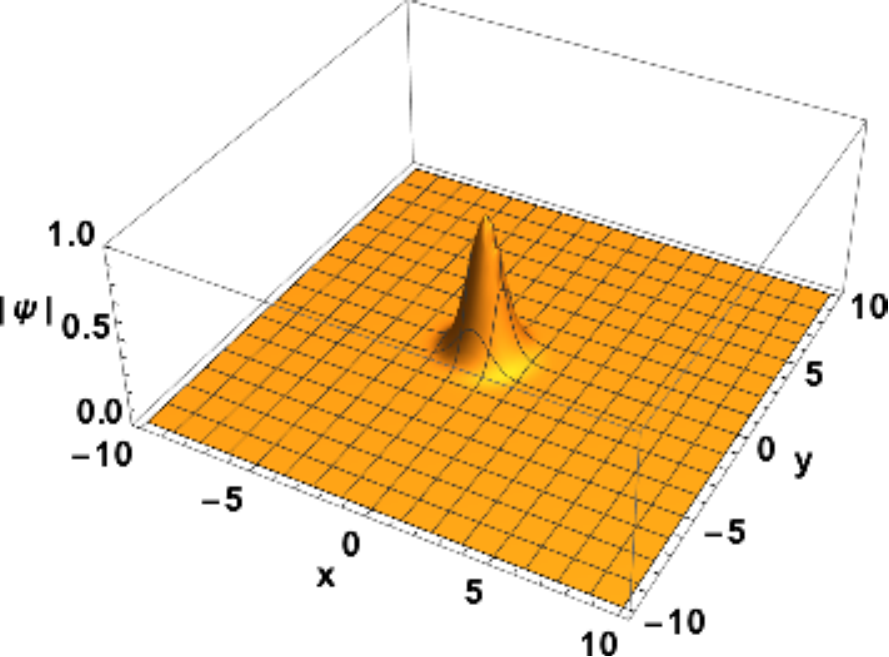}
\includegraphics[width=0.25\textwidth, clip=true]{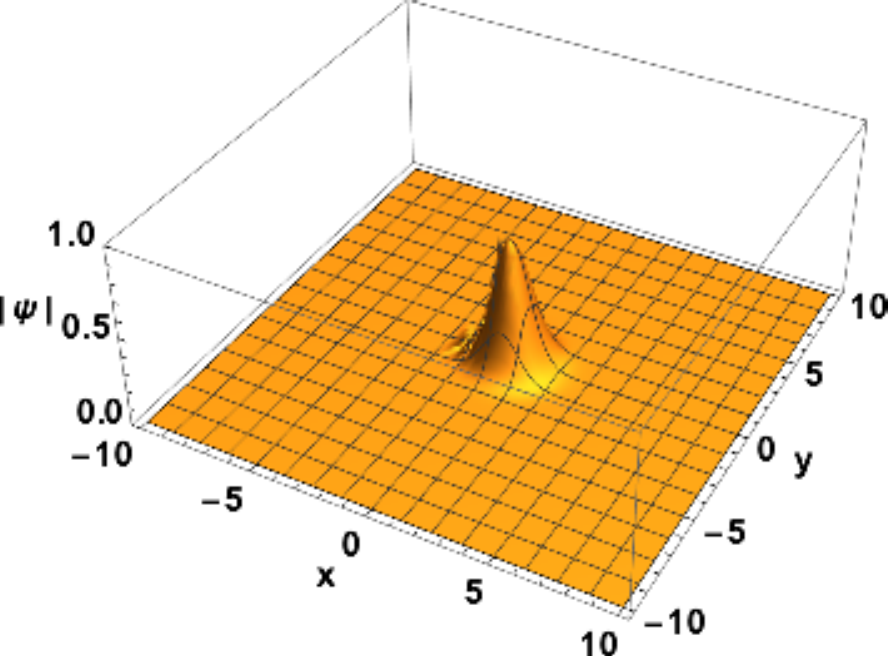}
\includegraphics[width=0.25\textwidth, clip=true]{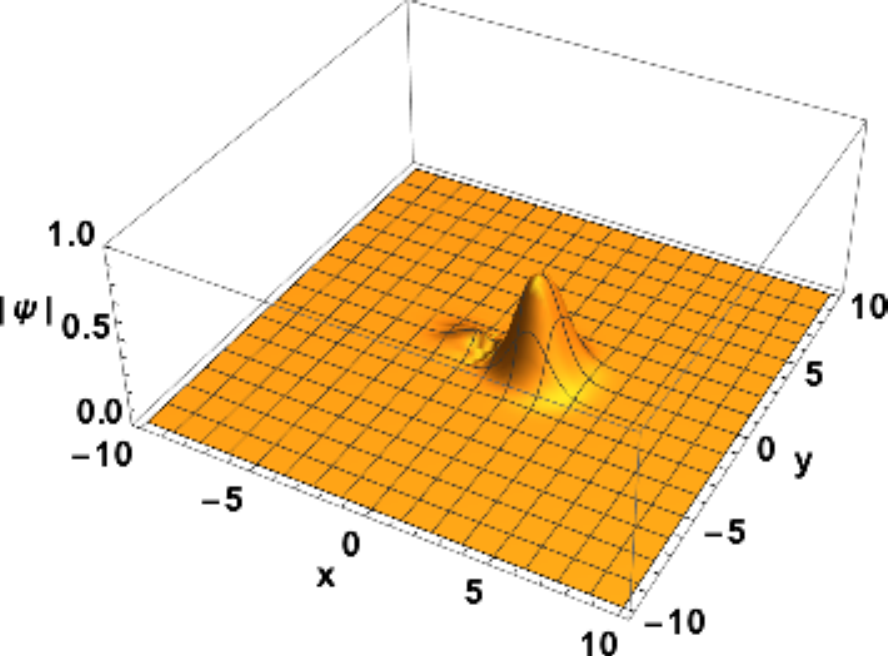}
\includegraphics[width=0.25\textwidth, clip=true]{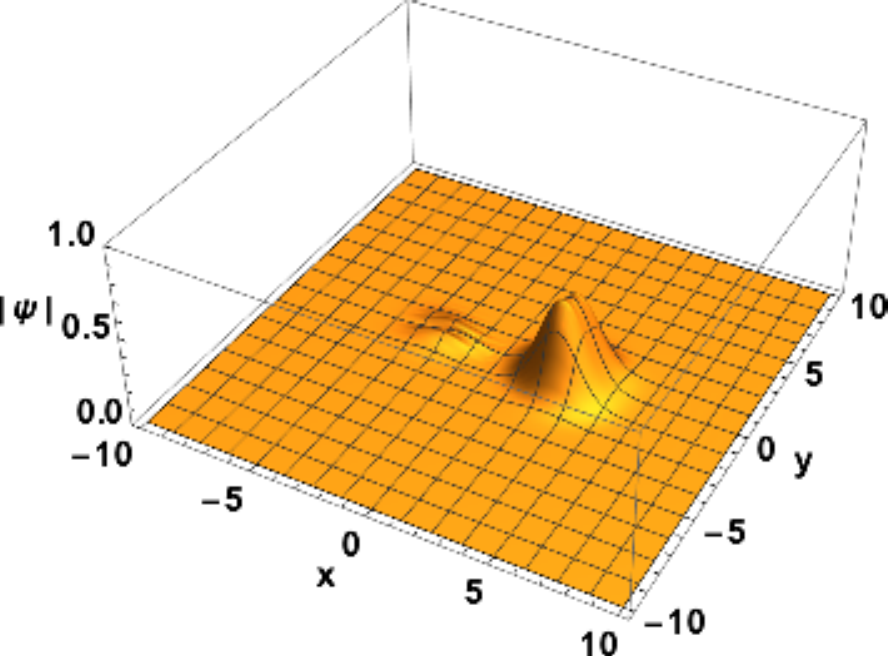}
\includegraphics[width=0.25\textwidth, clip=true]{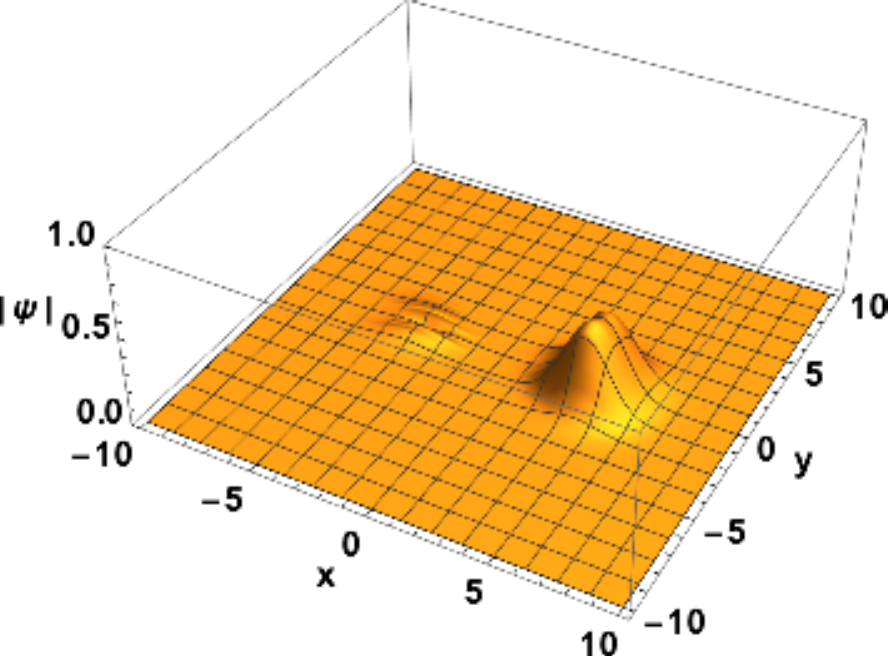}
\caption{Same as in Fig. 1 but for a combination of a repulsive Coulomb potential and an attractive Yukawa potential with equal strengths, and the same time sequence of snapshots. The third panel (top-right) already shows some difference with respect to the corresponding panel in Fig. 1, with a more uniform wave packet. The differences become more evident in the comparison of the last three panels of the two figures. The wave packet is slowed down in the presence of the Yukawa term, and there is reduced back-reflection.}
\label{Fig2}
\end{figure*}
Second, even for relativistic collisions we can consider electron-electron interactions are approximated by their static Coulomb potential, the magnetic field generated by two counter propagating beams made of same-charge particles being approximately zero. This also implies that, apart from the Lorentz boost of the cross-section, we do not expect relativistic considerations to play a major role in the interaction dynamics.
Finally, we neglect the effect of the electric charge on the horizon. The Reissner-Nordstr\"om radius is indeed negligible with respect to the Schwarzschild radius in the case of strong gravity and  collisions occurring above $\sqrt{s} \sim$ 30 GeV. 

\begin{figure*}[t]
\includegraphics[width=0.40\textwidth, clip=true]{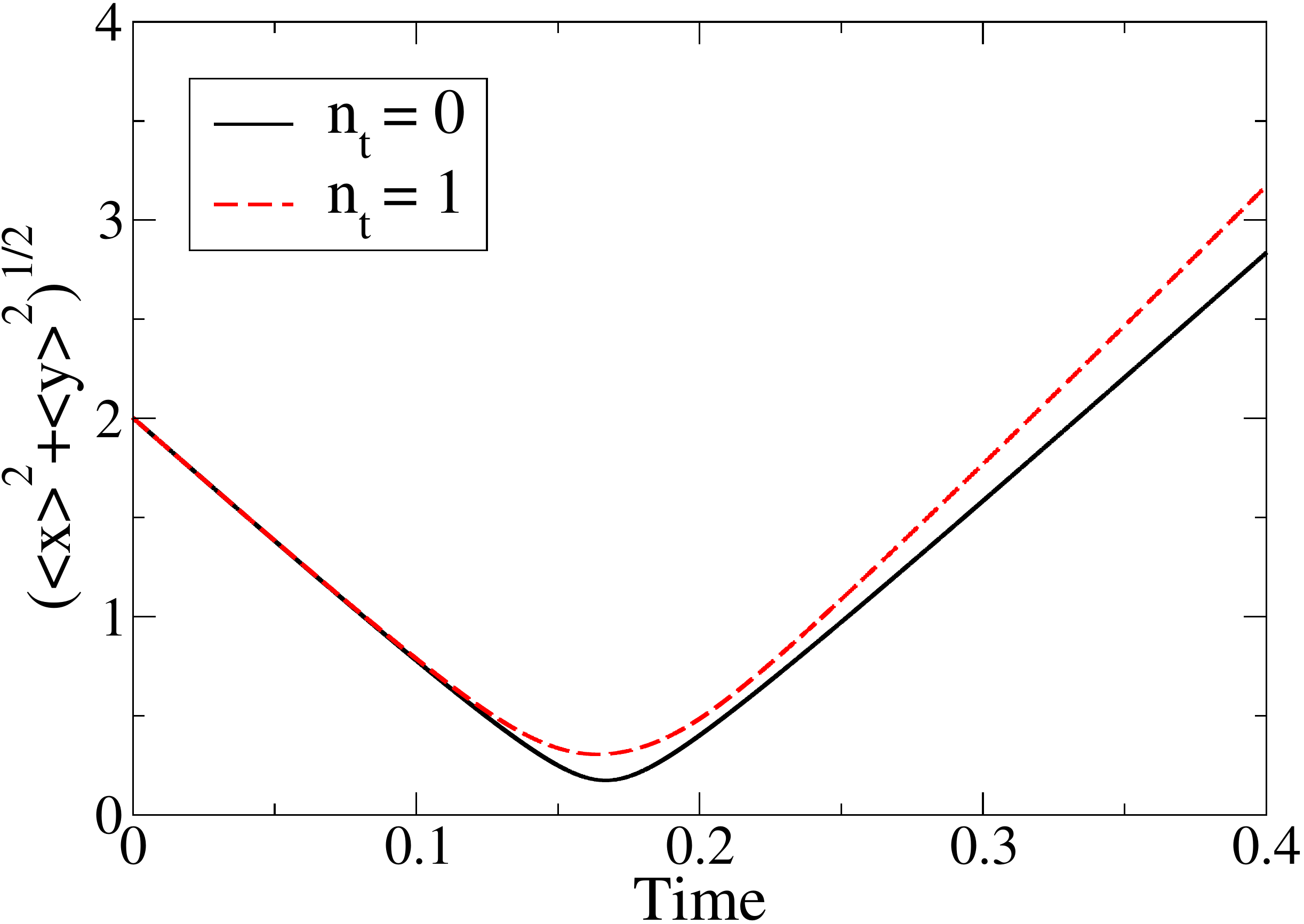}
\includegraphics[width=0.40\textwidth, clip=true]{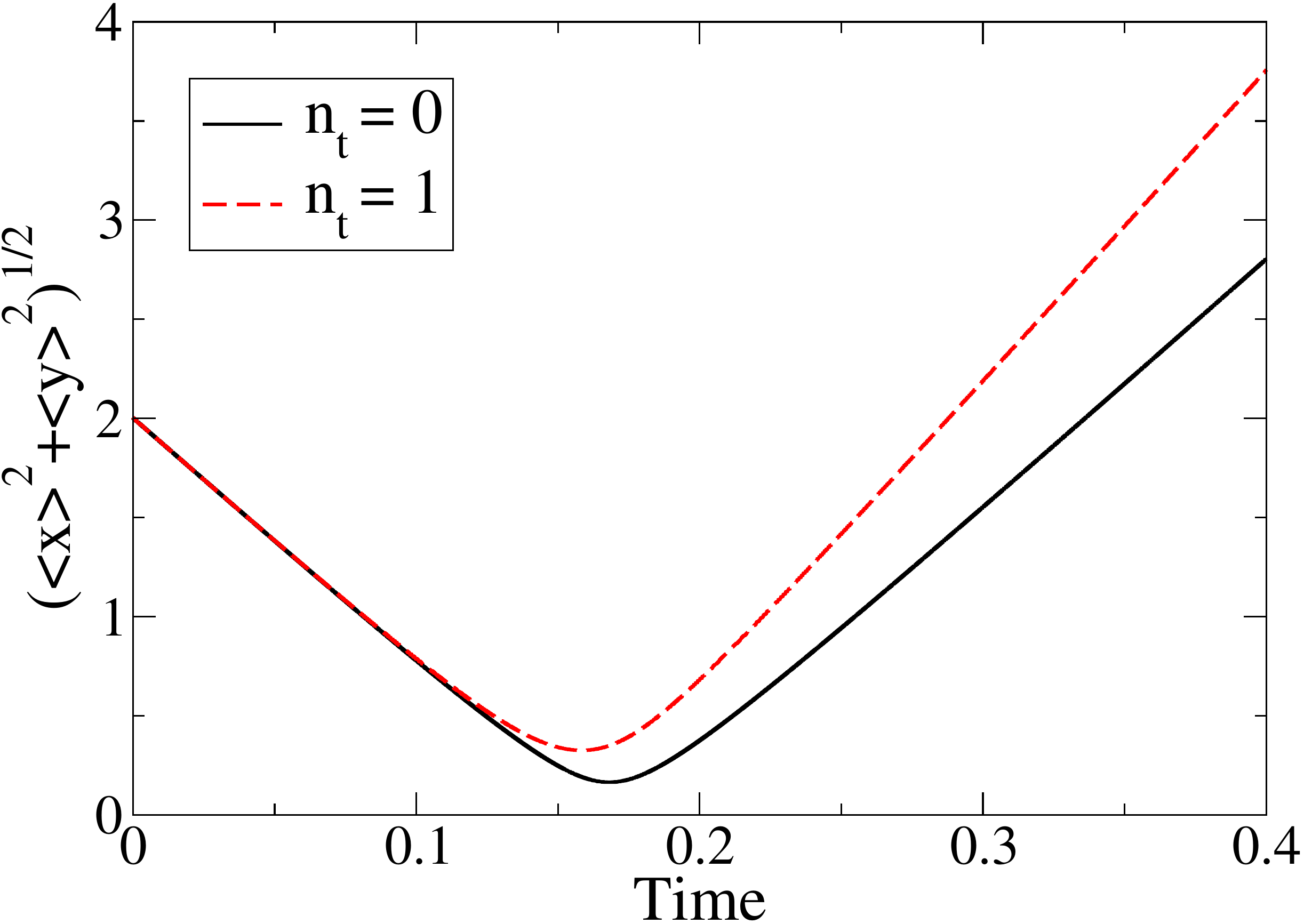}
\includegraphics[width=0.40\textwidth, clip=true]{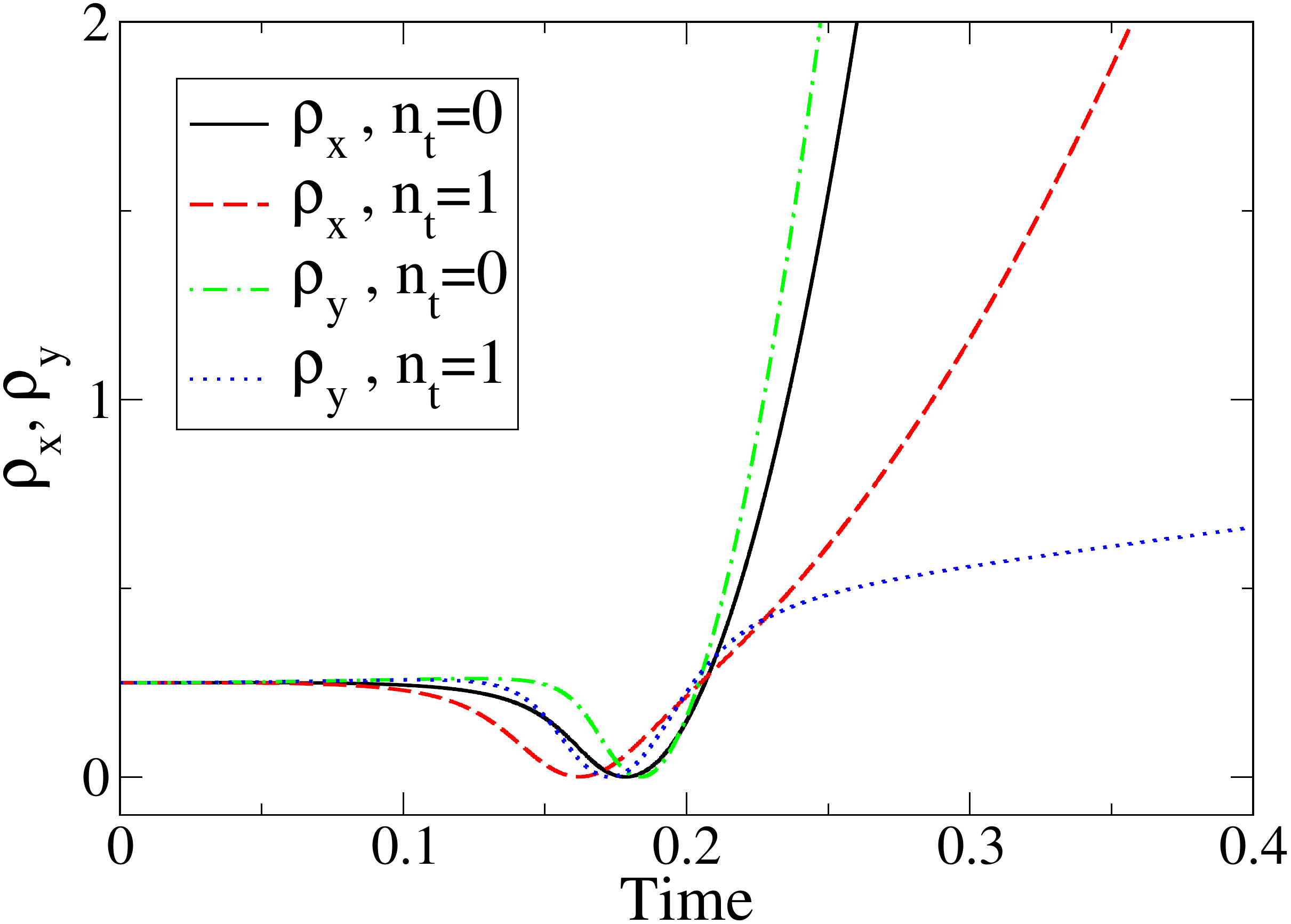}
\includegraphics[width=0.40\textwidth, clip=true]{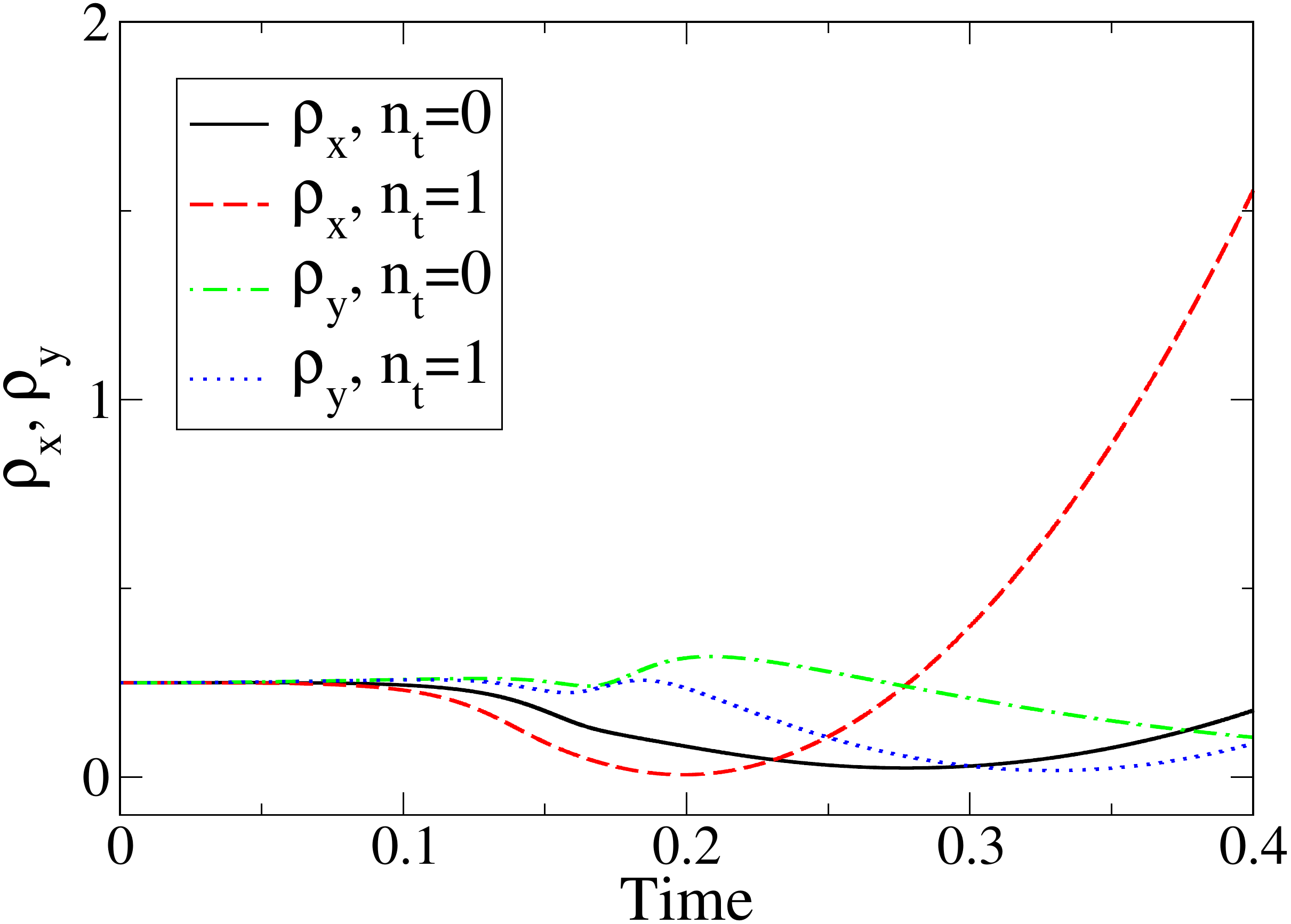}
\caption{Particle scattering through the Extended Gaussian Dynamics approximation. Plots of the average 
position in the $x-y$ plane (top left), and the Gaussian variances in the $x$ and $y$ plane (bottom left) for scattering 
from a repulsive Coulomb potential. On the right plots, same quantities but in the case of an attractive Yukawa  
potential added to the repulsive Coulomb potential. The integer $n_t$ denotes the order of truncation of the series present in Equations (4-7), such that $n_t = 0$ represent the purely classical dynamics, and $n_t=1$ the first order quantum corrections.}
\label{Fig3}
\end{figure*}

A first approach consists of numerically integrating the Schr\"odinger equation in a two-dimensional setting. 
One-dimensional problems do not capture the physics of the collisions as they miss the vast majority of 
events occurring at moderate, non-zero impact parameters. The full three-dimensional setting is obviously 
superfluous due to the cylindrical symmetry around the direction of the two beams. Furthermore, the problem of 
two colliding electrons interacting with central forces is transferable to the one of a single electron with reduced 
mass half its physical mass in the presence of a static potential equivalent to the one between the two electrons, plus the attractive correction due to usual gravitation (completely negligible in a realistic case). This corresponds to a potential energy $U_C=[-e^2/(4\pi \epsilon_0)+G_N m^2]/r=\alpha_C/r$, with $\epsilon_0$ the dielectric permittivity of vacuum.
In Fig. 1 we present the modulus of the wave function $|\psi(x,y,t)|$ at various instants of time numerically evaluated from an initial condition of a Gaussian wave packet impinging on a static 
repulsive Coulomb potential.  The wave packet spreads and starts to experience interference effects 
while approaching the singularity of the potential.  Notice that at later time the wave function ``fragments" into various components for 
small impact parameters. Analogous simulations for larger impact parameters confirm that 
the wave packet roughly maintains a Gaussian shape while following an average trajectory being repelled away from 
the scattering center. In presence of an additional attractive Yukawa potential representing the short-range component in Equation (\ref{Yukawa}) , the dynamics becomes more complex, as 
seen in Fig. 2.  If the strength of the Yukawa potential is small, as expected there are small deviations from the purely 
Coulomb scattering, at least for wave packets with energy small enough to avoid penetration into the scattering center in 
which the Yukawa potential is significantly strong. In this case the bulk of the dynamics can be described in terms of an 
effective potential which is still Coulomb but of smaller strength, implying smaller deflections of the electron trajectory.
However, if the Yukawa potential is large enough and/or in the presence of a large Yukawa range, the dynamics is significantly 
affected, as a larger component of the wave function propagates through the central potential due to reduced repulsion. 
At smaller initial momentum, the wave function tends also to be more localized momentarily in the local minimum of the 
potential at the scattering center, delaying its exit. 
The numerical method has, as usual, the drawback of being time-consuming, on top of the lack of accuracy due to the 
numerical approximations and the presence of finite steps in time and space, and of finite boundaries.

An alternative which has rather different approximations, and therefore allows for a cross-check of these numerical results, is provided 
by the so-called Extended Gaussian Dynamics (EGD). This belongs to the family of semiclassical methods 
originated by Ehrenfest \cite{Ehrenfest}. In the simplest case of a particle described in one-dimension, the Heisenberg equations 
of motions are averaged with a suitable Taylor  expansion of the potential energy $V(x)$, obtaining an infinite hierarchy of equations, 
each corresponding to higher order moments of position and momentum \cite{Hanggi,Ballentine}:
\begin{eqnarray}
\frac{d\langle x \rangle}{dt} &=& \frac{\langle p \rangle}{m},
\\ 
\frac{d\langle p \rangle}{dt} &=& - \sum_{n=0}^{\infty} \frac{1}{n!} V^{(n+1)}(\langle x \rangle) \langle \Delta \hat{x}^n \rangle,
\end{eqnarray} 
where $V^{(n+1)}=\partial^n V/\partial x^n$. The EGD consists in truncating this infinite set of equations by demanding that 
at any given time the state has a Gaussian form, thereby characterized by only two cumulants \cite{Robertson,Schroedinger}:

\begin{eqnarray}
\frac{d \langle x \rangle}{dt} &=& \frac{\langle p \rangle}{m},
\\ 
\frac{d \langle p \rangle}{dt} &=& - \sum_{n=0}^{\infty} \frac{1}{n! 2^n} V^{(2n+1)}(\langle x \rangle) \rho^{2n},
\\
\frac{d \rho}{dt} &=& \frac{\Pi}{m},
\\
\frac{d \Pi}{dt} &=& \frac{\hbar^2}{4m\rho^3}- \sum_{n=0}^{\infty} \frac{1}{n! 2^n} V^{(2n+2)}(\langle x \rangle) \rho^{2n+1},
\end{eqnarray} 
Here the variable $\rho$ is defined such that the even cumulants have the expression $\Delta \hat{x}^{2n}= \rho^{2n} 2n!/(2^n n!)$, 
starting from $n=0$, and the variable $\Pi=\langle \Delta \hat{x} \Delta \hat{p}+ \Delta \hat{p} \Delta \hat{x}/(2\rho)$. In this way the dynamics is described in term of the centroids of the Gaussian in ordinary phase space, and of an associated ``fluctuational phase space" in which the variances of position and momentum evolve in time.

\begin{figure*}[t]
\includegraphics[width=0.45\textwidth, clip=true]{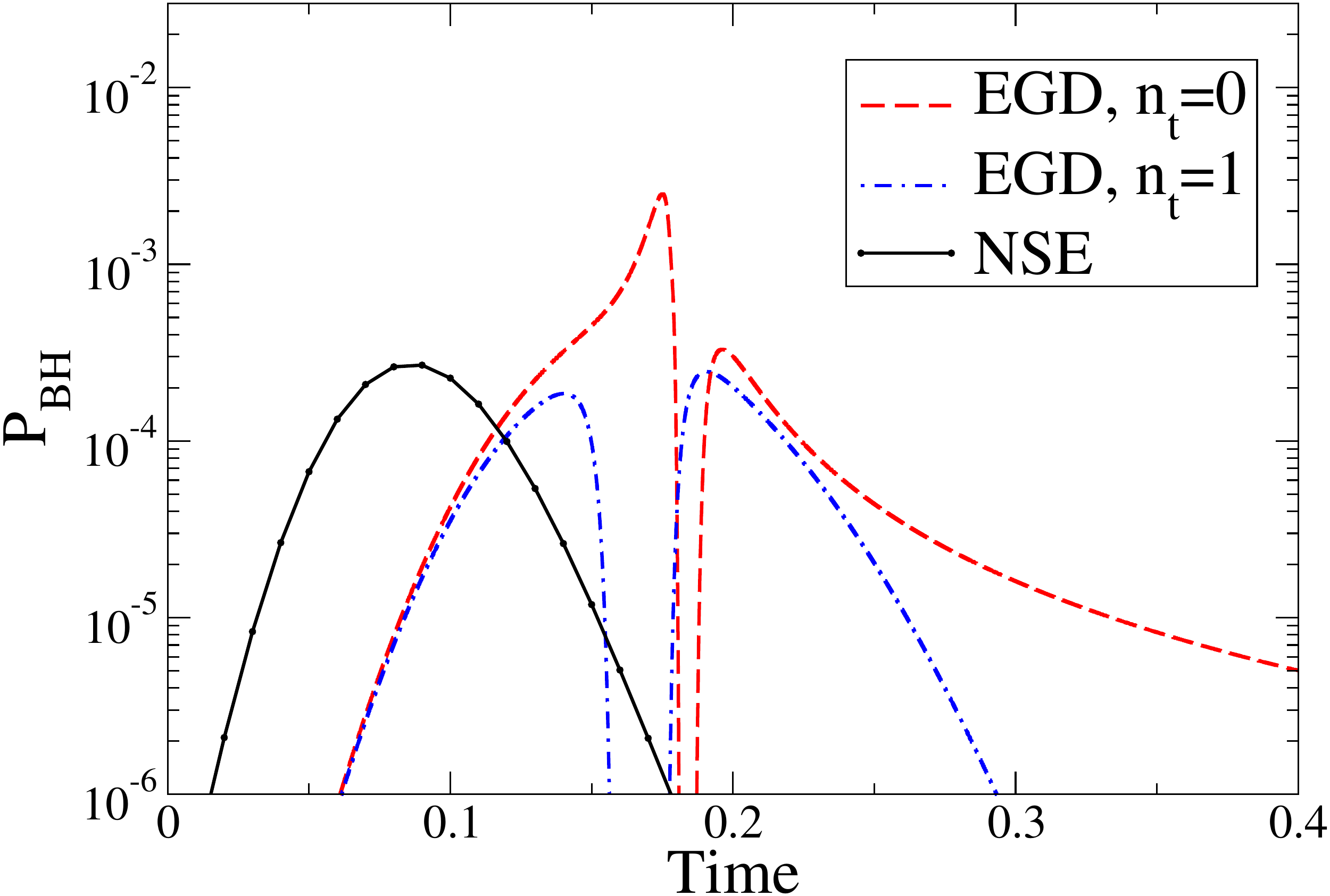}
\includegraphics[width=0.45\textwidth, clip=true]{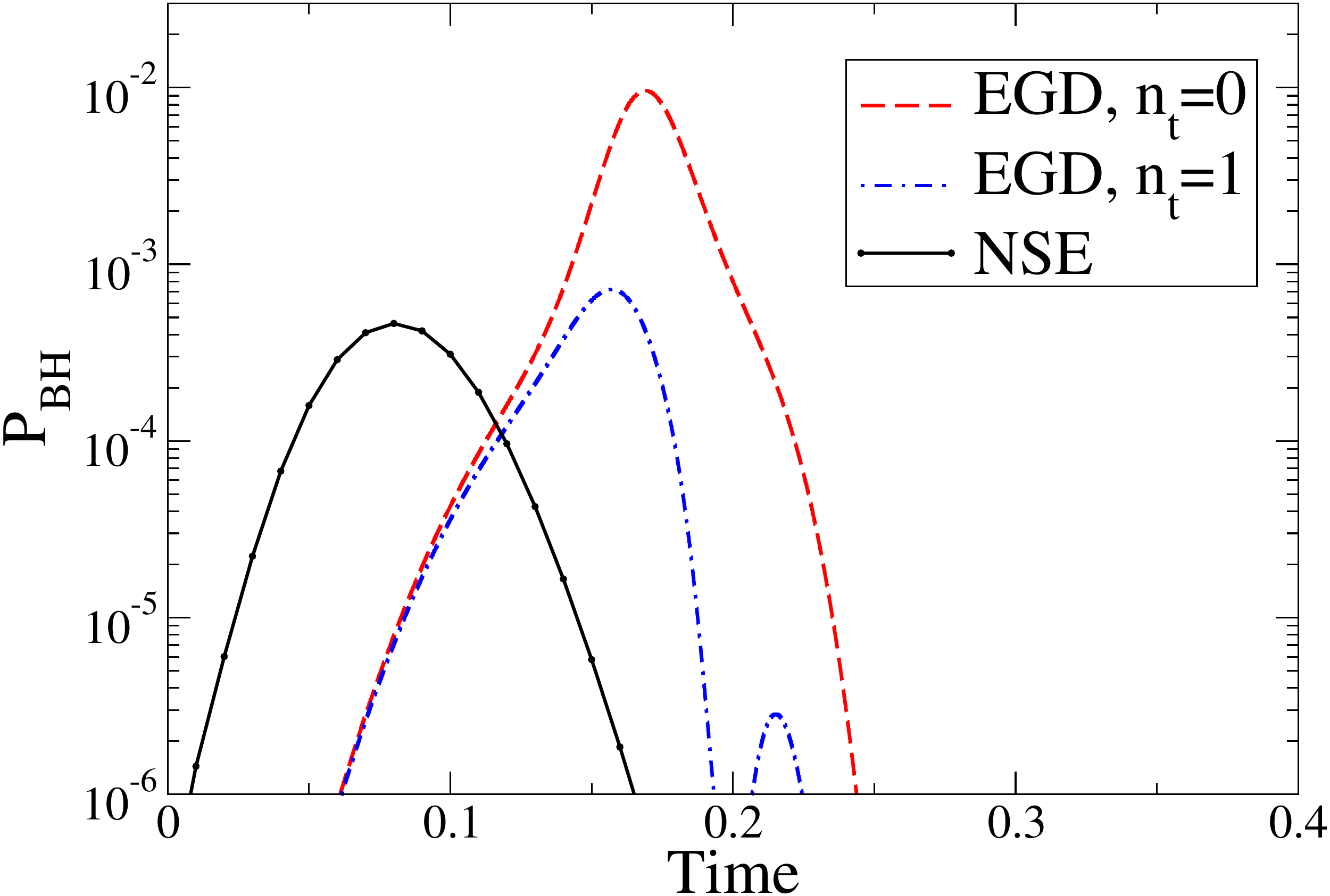}
\caption{Black hole production probability for electron-electron collisions versus time evaluated through 
numerical simulations of the Schr\"odinger equation (NSE, solid line) and the Extended Gaussian Dynamics (EGD) at orders of approximations zero (dashed line) and one (dot-dashed line). The left plot is for a repulsive Coulomb 
potential, the right plot includes also an attractive Yukawa potential, for the same parameters as in Figure 3. Notice the more complex dynamics for the EGD cases, due to the breathing of the Gaussian wavefunction occuring especially in correspondence of the region of minimum approach distance, around a time of 0.17.}
\label{Fig4}
\end{figure*}

This technique has been extensively used in chemical physics \cite{Heller} and in atomic trapping \cite{Choi}, but the constraint of a Gaussian shape is obviously rather strong. For instance, shortcomings of its application to potentials in which tunneling phenomena occur, with consequent creation of delocalized wave functions without a classical counterpart, has been discussed in \cite{Hasegawa}. In our situation, we will see that the presence of interference terms during the collision will appear in a rather approximate and peculiar form. In our specific relativistic setting, the only modification occurs in Eqs. (4) and (6), as the relativistic factor $\gamma$ will appear in the denominator on the right hand side, {\it i.e.} $m \rightarrow \gamma m$.

To provide a benchmark for the EGD method in a simple situation, we can consider the motion of a harmonic oscillator of 
mass $m$ and angular frequency $\omega$.  In this case Eqs. (4) and (5) assume a simple form, with the right hand sides equal,  
respectively, to $-m \omega^2 \langle x \rangle$ and $\hbar^2/(4 m \rho^3) - m \omega^2 \rho$. The motion is harmonic 
in the  centroid space ($\langle x \rangle, \langle p \rangle$), and the variance $\rho=\Delta \hat{x}$ has a motion resulting from 
the interplay between a harmonic-like force dominating at large $\rho$ and a strong repulsive effective force at small $\rho$. 
The latter term enforces the uncertainty principle making sure than $\rho$ cannot become zero, a sort of ``centrifugal" term which 
obviously goes to zero in the classical limit. The motions in the ($\langle x \rangle, \langle p \rangle$) and ($\rho, \Pi)$ variables 
are decoupled, and the centroid moves according to the classical Hamilton equations, while the $\rho$ variable may oscillate 
around its minimum value, {\it i.e.} the Gaussian state is ``breathing", unless it is a minimum uncertainty Gaussian state. 

In a general case, the equations of motion for the centroid and for the fluctuational space are coupled, leading to quantum 
corrections to the classical trajectory. The presence of an external potential in general affects the dynamics of the positional variance. 
Power-law potentials like a linear one are not enough to decrease the positional variance during the dynamics, while a quadratic 
potential results in a stationary (or periodically breathing) positional variance. Instead, potentials steeper than a parabolic one, like cubic or quartic potentials, or potentials depending on inverse distance as the Coulomb one, can momentarily decrease the positional variance. 
In our toy model, we have chosen the relevant parameters to be $\alpha_C = 10$, $\lambda_H = 0.1$, $\alpha_H = 15$, $R_S = 10^{-2}$, with an initial spreading of 0.25 in the $x$-direction, and an impact parameter of 0.1 (all quantities in Figs. 1-4 are in arbitrary, computer-friendly, units). These conditions are taken to mimic the realistic case in which (a) the Schwarzschild radius under Newtonian gravity $R_S$ is much smaller then the spread of the particle, (b) the Coulomb potential and the Yukawa potential at its maximum strength are close in amplitude, and (c) the range of the Yukawa potential is much smaller than the positional spread of the particle. Figure 3 shows that the presence of the Yukawa term significantly reduces the variance of the particles at later time, as expected due to its attractive character. We notice that with the Coulomb potential, in the $n_t = 0$ case  $\rho_x$ grows faster than in the $n_t = 1$ case, whereas the opposite happens with the Coulomb plus attractive Yukawa potential. This is because the inclusion of higher order terms in equation (7) makes the effect of the Yukawa potential on the variance much stronger than that with lower orders of expansion, thus canceling out the effect of the Coulomb term on the positional variance. Due to Coulomb repulsion, the variance of the particle decreases first, while in the presence of a substantial Yukawa term the opposite occurs. Figs. 1-2 already show that the wave function is more strongly reflected in the Coulomb potential than in the Coulomb plus attractive Yukawa potential, resulting in a larger variance in the Gaussian approximation in the former case. This effect, confirmed in the behavior of the variances shown in Fig. 3, is of relevance for the considerations reported in the following section. 

\section{Black hole production probability}

In this section we define a probability for black hole production based on the hoop conjecture in the presence of interactions, and estimate its magnitude using the numerical and EGD methods previously discussed.  Consider the wave functions of the two colliding particles $\psi(x,t)$, $\psi'(x',t)$. At time $t$, the probability of black hole formation is the probability that the two particles are located within the time-dependent Schwarzschild radius $\tilde{R}_S$. The differential probability that a black hole is formed if the a particle is in the position interval $(x, x+dx)$ and the other particle is located nearby within a Schwarzschild radius 
, {\it i.e.} it lies in the interval $(x'-\tilde{R}_s, x'+\tilde{R}_s$), is

\begin{equation}
dP_{BH}(t) = \int_{x}^{x+dx}|\psi(x',t)|^2 dx' 
\int_{x'-\tilde{R}_S}^{x'+\tilde{R}_S}|\psi'(x'',t)|^2 dx'',
\end{equation}
which in the realistic case of small $\tilde{R}_s$ becomes 

\begin{equation}
dP_{BH} = 2\tilde{R}_S(t)\int_{x}^{x+dx}|\psi(x',t)|^2  |\psi'(x',t)|^2 dx'.
\end{equation}

The total probability for black hole formation is obtained by integrating over 
all possible values of $x$

\begin{equation}
P_{BH}(t) = 2 \tilde{R}_S(t)\int_{-\infty}^{\infty}|\psi(x',t)|^2  |\psi'(x',t)|^2 dx'.
\label{EquationPBH1D}    
\end{equation}

The EGD lends itself to a simple evaluation of this probability, provided that the assumption of a  general Gaussian form for the wave function holds at the relevant times, in this case during the closest approach in the collision. If the general Gaussian wave function is characterized by its centroid $x_0(t)$, its wave vector $k_0(t)$ and its positional spread $\rho_x(t)$, we will have

\begin{figure*}[t]
\includegraphics[width=0.40\textwidth, clip=true]{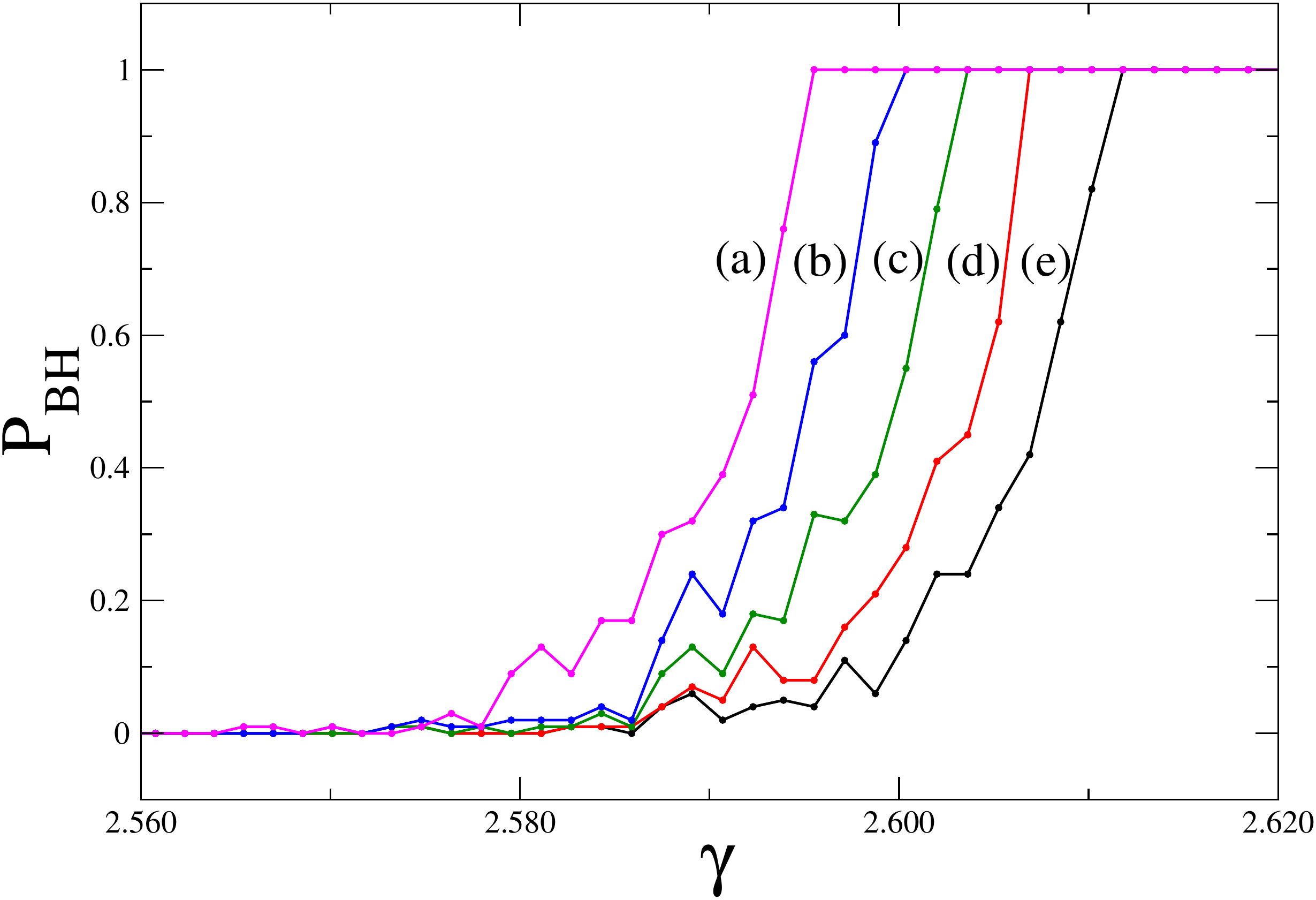}
\includegraphics[width=0.40\textwidth, clip=true]{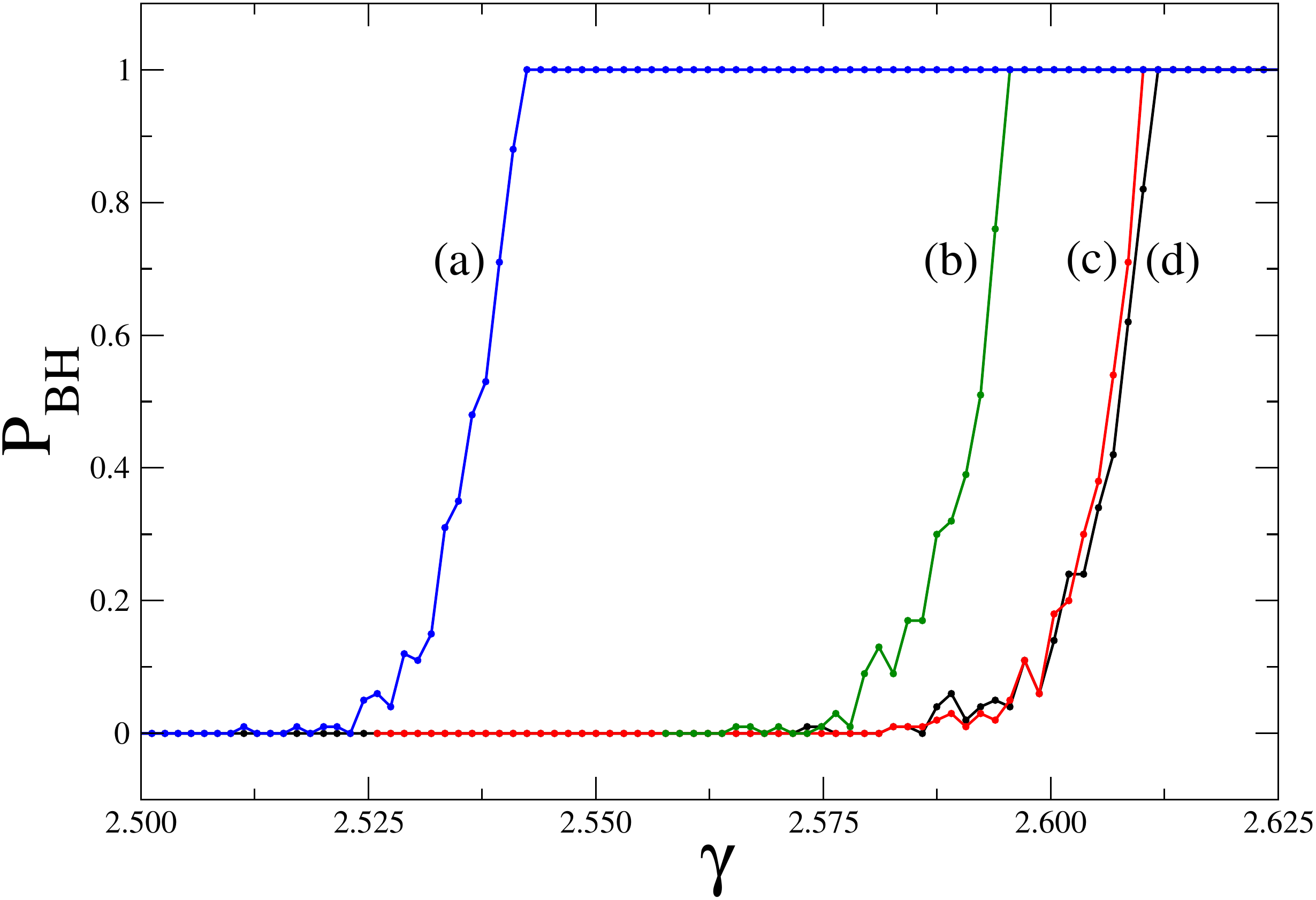}
\caption{Black hole production probability for electron-electron collisions versus relativistic factor $\gamma$ evaluated with the Extended Gaussian Dynamics for a repulsive Coulomb potential and attractive Yukawa potential added to the Coulomb strength of various relative strengths and ranges. The left plot shows the cases for an attractive coupling equal to (a) $\alpha_H/\alpha_C=0.020$, (b) $\alpha_H/\alpha_C=0.015$, (c) $\alpha_H/\alpha_C=0.010$, (d) $\alpha_H/\alpha_C=0.005$, all with range $\lambda_H=1$  (in arbitrary units). The case of a purely Coulomb 
repulsion (e) is also shown. The right plot shows the cases for a constant attractive coupling $\alpha_H/\alpha_C= 0.020$ with different ranges (a) $\lambda_H=2$, (b) $\lambda_H=1$, and (c) $\lambda_H=0.5$, while case (d) again is the pure Coulomb repulsion for comparison. The impact parameters are generated by a Gaussian distribution centered around zero and with variance $10^2$, and the black hole probability is evaluated on a sample of $10^2$ particles for each beam.} 
\label{Fig5}
\end{figure*}

\begin{equation}
\psi(x,t) = \frac{1}{\sqrt{\rho_x\sqrt{\pi}}} \exp\left\{-\frac{[x-x_0(t)]^2}{2\rho_x(t)^2}+{ik_0(t)x}\right\},
\end{equation}
while the other colliding particle will have the opposite sign for the wave vector.  
Therefore, Eq.  (\ref{EquationPBH1D}) will become 

\begin{equation}
P_{BH}(t) = \frac{2\sqrt{2}}{\pi} \frac{\tilde{R}_S(t)}{\rho_x(t)} \exp\left[-\frac{2x_0(t)^2}{\rho_x(t)^2}\right],
\end{equation}
which can be easily generalized to the two-dimensional case

\begin{equation}
P_{BH}(t) = \frac{8\tilde{R}_S(t)^2}{\pi \rho_x(t)\rho_y(t)}\exp{\left[-2\left(\frac{x_0(t)^2}{\rho_x(t)^2}+\frac{y_0(t)^2}{\rho_y(t)^2}\right)\right]}.
\label{EquationPBH2D}
\end{equation}

Equation (\ref{EquationPBH2D}) allows us to replace complex numerical simulations with a relatively simple formula containing time-dependent quantities, as emphasized by the explicit time-dependence for each relevant quantity. This avoids numerical issues related to the presence of quantities, such as the coupling strength of the effective interaction, or the relevant length scales that can vary over several orders of magnitude. 
Equation (\ref{EquationPBH2D}) also shows that the probability of black hole formation  decreases exponentially with the distance between the particles, as intuitively expected, although each component of the distance gets weighted by the corresponding positional variance. 
The dependence upon the latter quantity is more subtle. In general, the positional variance increases from the initial state with the possibility of oscillations due to the breathing of the wave function in the presence of the potential. This may lead to a non-trivial interplay because, on one hand, the positional variances appear with inverse dependence in the denominator of the right hand side in Eq. (\ref{EquationPBH2D}) but, on the other hand, they also appear in the denominator of the exponential term of the same equation. The first term tends to decrease the probability for forming black holes as the wave function of the particle initially spreads out. However, the second, exponential term will yield a smaller suppression of the probability in the presence of the same spreading. It is then possible to expect, for a judicious choice of the initial parameters, a positional variance for which the formation of black holes is maximized, due to the fact that the interaction between the particles will tend to decrease their positional variance. In Fig. \ref{Fig4} we show an example of probability for black hole formation versus time, for the same average trajectories and the related variances discussed in Fig. \ref{Fig3}. This allows for a direct comparison between the  EGD method and the results of the numerical integration of the Schr\"odinger equation.
The presence of breathing effects is evidenced in the EGD approach, and becomes more pronounced as more terms in the expansion are included. As expected, the numerical solution of the Schr\"odinger equation and the EGD method do not provide the same probabilities and, in particular, it is evident that the peak probability occurs at a later time for the EGD method, as any non-Gaussian precursor is not included in the related approximation. However, the first order  expansion of the EGD equations ($n_t=1$) is much closer to the numerical solution. Comparing the two plots, it is evident with either method that the boosting of the effective Schwarzschild radius due to the Higgs component enhances the probability of black hole production. 

\section{Experimental considerations}

We discuss in this section possible ways to observe the expected enhancement of the probability for black hole production in electron-electron collisions. 
The main idea, differing from the hadronic case analyzed at the LHC, is that one should focus on precision measurements in the elastic scattering channel alone. The absence of the annihilation $s$-channel, present for electron-positron collisions, make the overall analysis more clean. 

\begin{figure*}[t]
\includegraphics[width=0.40\textwidth, clip=true]{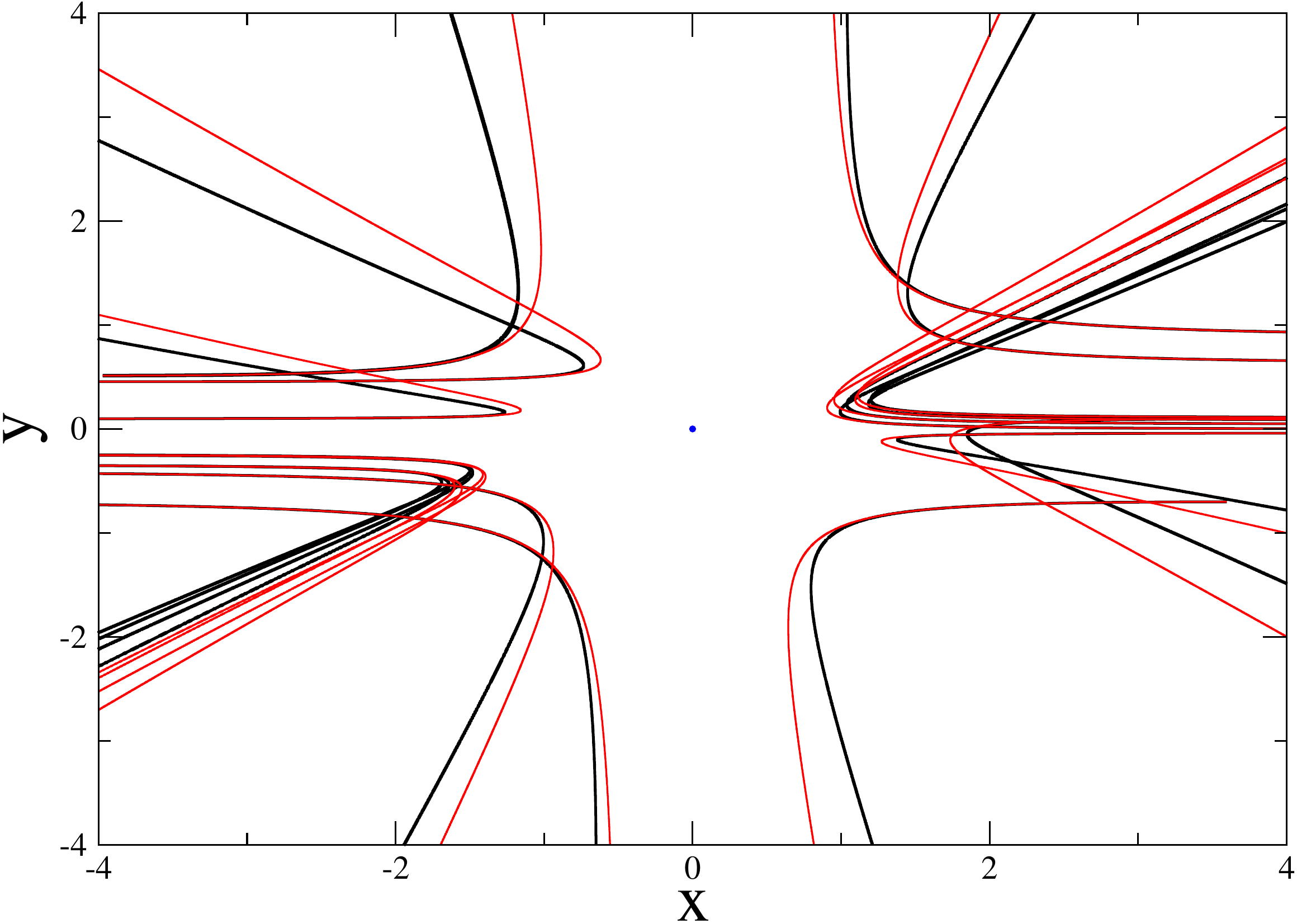}
\includegraphics[width=0.40\textwidth, clip=true]{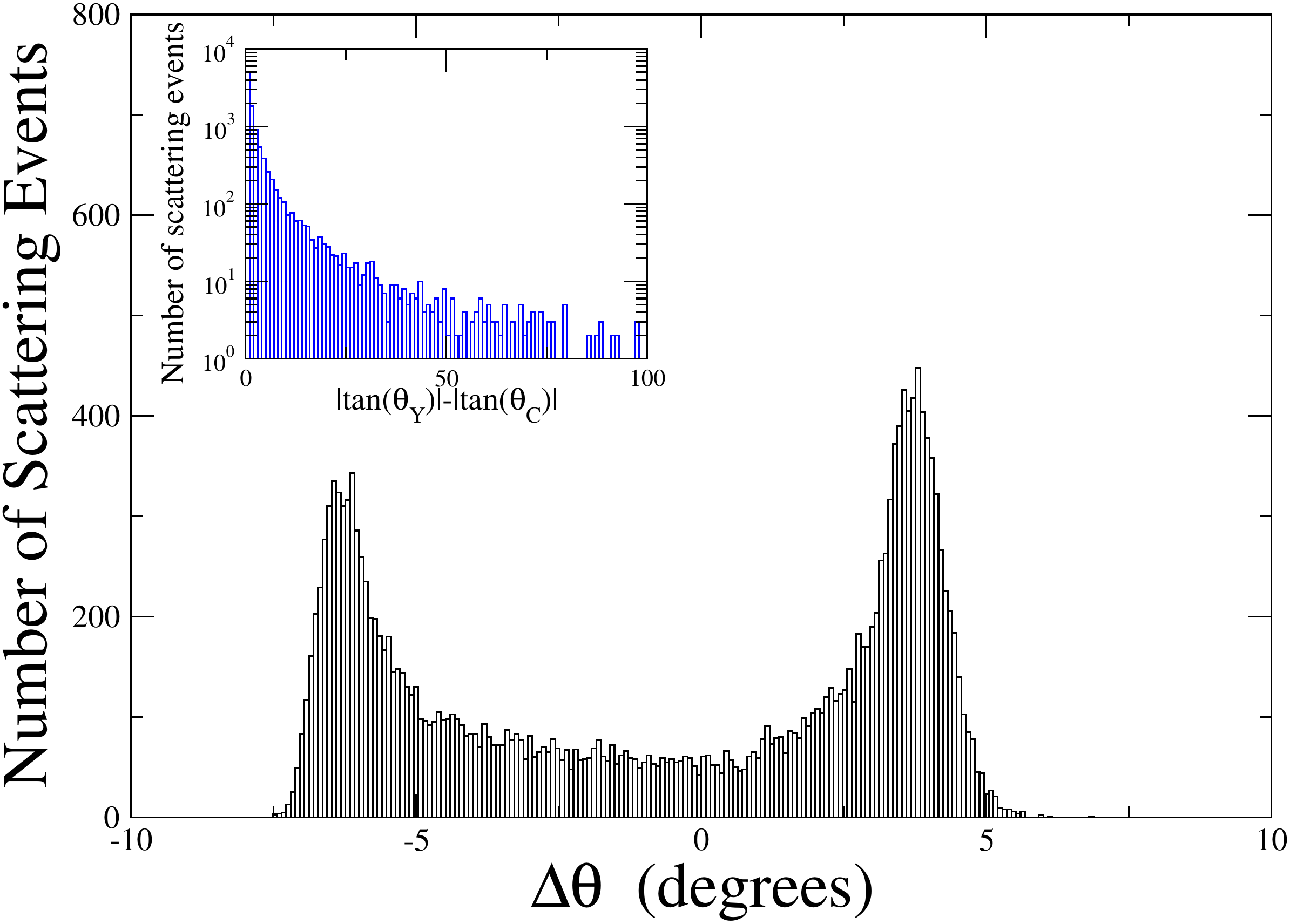}
\caption{Impact of the Yukawa potential on the kinematics of electron-electron collisions. On the left, a close-up of the collision region for a sample of nine collisions in the case of purely Coulomb interaction (black, thicker lines) and the same collisions with the Coulomb interaction plus a Yukawa attractive interaction with strength ten times the Coulomb interaction and range of $\lambda=0.5$ (red, thinner lines). The geometrical center of the collisions is evidenced by a dot. Notice that the presence of the Yukawa interaction leads, with respect to the corresponding trajectories in the presence of the pure Coulomb repulsion, to softer deflection angles and a smaller distance of approach between the two electrons. On the right, the distribution of the change in the scattering angle, measured with respect to the x-axis, due to the presence of the Yukawa potential, $\Delta \theta=\theta_Y-\theta_C$, is evaluated in the case of a simulation with $10^4$ collisions for the same parameters used in the plot on the left. The presence of negative values is due to trajectories, as in the top-left quadrant in the left panel, in which the Yukawa trajectory occurs at a smaller angle with respect to the Coulomb one. A more representative plot is depicted in the inset, which shows the distribution of the difference between the absolute values of the tangents of the related angles, $|\tan\theta_Y|-|\tan\theta_C|$. 
The asymmetry present in the $\Delta \theta$ distribution originates from the convolution with the horizontal spreading of the initial conditions visible, 
for instance, in the left figure with a trajectory originating from a smaller distance from the center in the bottom-left section. }
\label{Fig6}
\end{figure*}

In principle, the cleanest signature of the black hole production is achieved by energy scanning. Due to the hoop conjecture, we do expect a sudden increase in the production probability when the two electrons come within a Schwarzschild radius of each other. Once this happens, any small increase in energy will make the black hole production more likely, both due to the Lorentz boost in the Schwarzschild radii and the smaller minimum distance between the electrons. Such a sudden phenomenon is shown in both panels of Fig. 5 already for electron-electron collisions without the expected Yukawa-like component due to the Higgs field, see black dots and curves in both plots. The presence of the Higgs field is manifested in a precocious black hole formation, at a threshold value of the relativistic $\gamma$ factor occurring progressively earlier for largest values of the Higgs coupling strength and its range. Note also the larger sensitivity of the black hole probability to the Yukawa range due to the exponential dependence of the potential on this parameter. 

Another observable affected by the presence of a Higgs potential is the angular distribution, as we expect both a softening of the hard deflections due to the Coulomb repulsion, as well as smaller distances of approach between the two electrons. This is evidenced again in the toy model discussed above in the left plot of Fig. 6, where a subset of nine electron-electron  collisions is shown with a random generation of initial conditions in the x-y plane. The presence of softer deflections is quantified in the right plot of Figure 6, where we present the distribution of the difference between the deflection angles $\Delta \theta=\theta_Y -\theta_C$, where $\theta_Y$ is the deflection angle of the trajectory including the Yukawa potential, and $\theta_C$ is the deflection angle of the corresponding trajectory ({\it i.e.} with the same initial conditions) with purely Coulomb repulsion. With this observable, one expects larger deflection angles for the added Yukawa case for electrons originating from the right beam, and smaller deflection  angles for electrons originating from the left beam. This induces a nearly symmetrical spreading in the $\Delta \theta$ variable, subjected to asymmetries due to the different initial conditions in the positions of each electron pair. In a realistic setting, tracking this observable implies precision determination of the rapidity of the electron trajectories, and comparison to the expected QED predictions for a variety of energies. The effect of the Higgs field should appear as a systematic shift of the deflection angle with increasing energy of the beams, see inset of Figure 6. 

In terms of absolute cross-sections, an order of magnitude estimate is obtained by considering the electron-electron elastic (M\"oller) scattering. By using an angular acceptance in between 0.05 and 3.09 radians, the total elastic cross section at $\sqrt{s}$=100 GeV is evaluated to be about 72 nb. For a minimum impact parameter of $10^{-18}$ m at the same energy, the probability for black hole production based on Eq. (\ref{EquationPBH2D}) yields $P_{BH} \simeq 7 \times 10^{-14}$, based on the assumption that the positional spread of each electron wave function in its rest frame is larger than the related Compton wavelength. The corresponding cross-section, due to the minuscule branching ratio, is therefore simply the product of the elastic cross-section and the black hole production probability, {\it i.e.} $\sigma_{BH} \sim 5 \times 10^{-12}$ nb, {\it i.e.} 12 orders of magnitude smaller than the probability expected on the basis of the hoop conjecture.
Notice that the probability for black hole production is proportional to the square of the Schwarzschild radius which, in turn, is proportional to the relativistic $\gamma$ factor. This probability scales as the square of the energy of the electrons, while the elastic cross section scales with its inverse. Therefore, the absolute cross section for black hole production is expected to be independent of energy, and higher energies will make the corresponding branching ratio higher, therefore increasing the signal-to-background ratio with respect to the uninteresting elastic collisions, until it reaches the unitarity limit, {\it i.e.} a branching ratio of 100 $\%$, as in the toy model example in Fig. 5. 

Notice that the cross-section can be also evaluated keeping in mind that the interaction potential is Coulomb-like at both large impact parameters (coinciding in the case with the Coulomb potential) and at very small impact parameters (smaller than the Yukawa range). 
In an intermediate region, the potential is adequately approximated by a linear and and a quadratic term, which satisfy the Ehrenfest theorem. Therefore, as far as elastic processes are considered, classical cross-sections ``a l\`a Rutherford" are adequate.  This also ensures a protection from genuine quantum effects and justifies the robustness of the EGD technique. In the context of a comprehensive analysis, checking for quantum corrections will be important \cite{Hsu,Cavaglia,Mureika}, including the probabilistic nature of the event horizon as already discussed, in a one-dimensional and non-interacting setting, in \cite{Casadio1,Casadio2,Casadio3}. 

In a concrete experimental proposal one should also take into account that, at the energies of interest, $Z^0$ exchange will superimpose to photon exchange \cite{Anthony}, but we do not expect the picture to change significantly at least below the electroweak breaking symmetry scale. Above this scale, we expect the unified running coupling constant to decrease, therefore partially mitigating the repulsive effect. 
In the regime of extremely high energies not yet available, the presence of asymptotic freedom for the unbroken non-Abelian gauge group of the electroweak model will allow for cross sections closer to the geometrical estimate based upon the hoop conjecture. Analogous considerations can be carried out for hadronic collisions, in which the simultaneous presence of color and electromagnetic interactions, at least for the quark degrees of freedom, complicates the analysis, as we plan to discuss in the future. Also, high-energy elastic scattering will be affected by radiative corrections due to the emission of hard photons via bremsstrahlung. 
This effect will be mitigated in a $\mu^+\mu^-$ collider. 

\section{Conclusions} 

Bounds to production of black holes in hadronic collisions as discussed in \cite{CMSBH1,CMSBH2,ATLASBH1,ATLASBH2} rely upon the hoop conjecture, which is of purely geometrical nature and therefore does not include the effect of interactions between the colliding particles. We have discussed the impact of interactions in the more controllable case of electron-electron collisions for a model of strong gravity living in four dimensions and including the effect of the Higgs field at the attometer scale. We have derived an analytical expression for the probability of black hole formation, and benchmarked its validity with numerical simulations and controlled semiclassical  approximation schemes. The presence of Higgs-induced strong gravity is not enough to significantly offset the suppression in cross-section due to the Coulomb repulsion. This result does not immediately impact the bounds discussed at the LHC since hard collisions between quarks will be mainly dominated by gluon exchange, of attractive character, at odds with the repulsive character of the same-charge electrons discussed here. Nevertheless, we envision possible bounds to black hole production in a more pristine environment once the new generation of leptonic colliders, either $e^+e^-$ or $\mu^+\mu^-$, will be operational, by simply converting the $e^+$ or $\mu^+/\mu^-$ beams to the opposite charge lepton, a rather simple modification at least for linear colliders \cite{Heusch}, complementing the already planned physics based on the  particle-antiparticle annihilation $s$-channel \cite{Craig}.  

Small impact parameters should be available through scattering between electrically neutral fundamental particles, such as high energy photons produced from $e^+e^-$ beams \cite{Bauer} . 
Also, a neutrino-neutrino collider seems in this regard an appealing, though unconventional, possibility. The extremely small mass and the corresponding small Schwarzschild radius even in strong gravity scenarios could be compensated by considering very high energy neutrinos, such as those expected from already planned muon colliders. The background to the process, consisting of particle production via neutrino annihilation into $Z^0$ bosons, is rather clean, but the strongest limitation will be set by the event statistics. 

\vspace{-0.1cm}

\acknowledgments

One of us (HQ) acknowledges financial support through the Junior Research Scholar Program at Dartmouth. We are grateful to Vincent P. Flynn and Terrence Kovacs for numerical help, and to Robert R. Caldwell, Leigh M. Norris, and Alexander R. H. Smith for useful comments.


\begin{thebibliography}{99}

\bibitem{Arkani} N. Arkani-Hamed, S. Dimopoulos, and G. R. Dvali, Phys. Lett. B 429 (1998) 263.

\bibitem{Antoniadis} I. Antoniadis, N. Arkani-Hamed, S. Dimopoulos, and G. R. Dvali, Phys.
Lett. B 436 (1998) 257.

\bibitem{Randall} L. Randall and R. Sundrum, Phys. Rev. Lett. 83 (1999) 3370.
  
\bibitem{Dvali1} G. Dvali and M. Redi, Phys. Rev. D 77 (2008) 045027.

\bibitem{Calmet} X. Calmet, S. D. Hsu, and D. Reeb, Phys. Rev. D 77 (2008) 125015.

\bibitem{Burinskii} A. Burinskii, Int. J. Mod. Phys. D 26 (2017) 1743022.

\bibitem{Thorne} K. S. Thorne, in {\sl Magic without Magic}, edited by J. R. Klauder 
(W.H. Freeman, San Francisco, 1972), pp. 231-258.
  
\bibitem{Giddings} S. B. Giddings and S. Thomas, Phys. Rev. D 65 (2002) 056010.  
  
\bibitem{Dimopoulos} S. Dimopoulos and G. Landsberg, Phys. Rev. Lett. 87 (2001) 161602.

\bibitem{Dvali2} G. Dvali, G. F. Giudice, C. Gomez, and A. Kehangias, JHEP 08 (2011) 108.

\bibitem{CMSBH1} CMS Collaboration, Phys. Lett. B 697 (2011) 434.

\bibitem{CMSBH2} CMS Collaboration, J. High Energ. Phys.  11 (2018) 042.

\bibitem{ATLASBH1} ATLAS Collaboration, Phys. Lett. B 760 (2016) 520.

\bibitem{ATLASBH2}  ATLAS Collaboration, J. High Energ. Phys. 03 (2016) 026.

\bibitem{Park} S. C. Park, Phys. Lett. B 701 (2011) 587.

\bibitem{Barkas} W. H. Barkas, R. W. Deutsch, F. C. Gilbert, and C. E. Violet, Phys. Rev. 86 (1952) 59.

\bibitem{Barber} W. C. Barber, G. K. O'Neill, B. Gittelman, and B. Richter, Phys. Rev. D 3 (1971) 2796.

\bibitem{ATLASHiggs} G. Aad {\it et al.} (ATLAS Collaboration), 
Phys. Lett. B 716 (2012) 1.

\bibitem{CMSHiggs} S. Chatrchiyan {\it et al.} (CMS Collaboration), 
Phys. Lett. B 716 (2012) 30. 

\bibitem{Dehnen1} H. Dehnen, H. Frommert, and F. Ghaboussi, Int. J. Theor. Phys. 29 (1990) 537.

\bibitem{Dehnen2} H. Dehnen, H. Frommert, and F. Ghaboussi, Int. J. Theor. Phys. 31 (1992) 109.

\bibitem{Onofrio1} R. Onofrio, Mod. Phys. Lett. 28 (2013) 1350022.

\bibitem{Onofrio2} R. Onofrio, EPL 104 (2013) 20002.

\bibitem{Onofrio3} R. Onofrio, Mod. Phys. Lett. 29 (2014) 1350187.

\bibitem{Brans} C. Brans and R. H. Dicke, Phys. Rev. 124 (1961) 925.

\bibitem{Hawking} S. W. Hawking, Commun. Math. Phys. 25 (1972) 167.

\bibitem{Scheel1} M. A. Scheel, S. L. Shapiro, and S. A. Teutolsky, Phys. Rev. D 51 (1995) 4208.

\bibitem{Scheel2} M. A. Scheel, S. L. Shapiro, and S. A. Teutolsky, Phys. Rev. D 51 (1995) 4236.

\bibitem{Sotiriou} T. Sotiriou and V. Faraoni, Phys. Rev. Lett. 108 (2012) 081103.

\bibitem{Ehrenfest} P. Ehrenfest, Zeit. Phys. 45 (1927) 455.

\bibitem{Hanggi} P. H\"anggi and P. Talkner, J. Stat. Phys. 22 (1980) 65.

\bibitem{Ballentine} L. E. Ballentine and S. M. McRae, Phys. Rev. A 58 (1998) 1799.

\bibitem{Robertson} H. P. Robertson, Phys. Rev. 34 (1929) 163.

\bibitem{Schroedinger} E. Schr\"odinger, Sitz. Preuss. Akad. Wissen. 14 (1930) 296.

\bibitem{Heller} E. J. Heller, J. Chem. Phys. 62 (1975) 1544.

\bibitem{Choi} S. Choi, R. Onofrio, and B. Sundaram, Phys. Rev. A 88 (2013) 053401.

\bibitem{Hasegawa} H. Hasegawa, Phys. Lett. A 378 (2014) 691.

\bibitem{Hsu} S. D. H. Hsu, Phys. Lett. B 555 (2003) 92.

\bibitem{Cavaglia} M. Cavagli\`a and S. Das, Class. Quantum Grav. 21 (2004) 4511.

\bibitem{Mureika} J. Mureika, P. Nicolini, and E. Spallucci, Phys. Rev. D 85 (2012) 106007.

\bibitem{Casadio1} R. Casadio and F. Scardigli, Eur. Phys. J C 74 (2014) 2685.

\bibitem{Casadio2} R. Casadio, O. Micu, and F. Scardigli, Phys. Lett. B 732 (2014) 105.

\bibitem{Casadio3} R. Casadio, R. T. Cavalcanti, A. Giugno, and J. Mureika, Phys. Lett. B 760 (2016) 36.

\bibitem{Anthony} P. L. Anthony, et al., (SLAC E158 Collaboration), Phys. Rev. Lett. 95 (2005) 081601.

\bibitem{Heusch} C. A. Heusch, Int. J. Mod. Phys. A 20 (2005) 7289.

\bibitem{Craig} N. Craig, {\sl A Case for Future Lepton Colliders},  arXiv: 1703:06079.

\bibitem{Bauer} D. Bauer, Int. J. Mod. Phys. A 11 (1996) 1637.


\end{thebibliography}
\end{document}